# Multi-element Germanium Detectors for Synchrotron Applications

Abdul K. Rumaiz[a], Anthony J. Kuczewski[a], Joseph Mead[a], Emerson Vernon[a], Donald Pinelli[a], Eric Dooryhee[a], Sanjit Ghose[a], Thomas Caswell[a], D. Peter Siddons[a] Antonino Miceli[b], Jonathan Baldwin[b], Jonathan Almer[b], John Okasinski[b], Orlando Quaranta[b], Russell Woods[b], Thomas Krings[c] and Stuart Stock[d]

[a] *Brookhaven National Laboratory,*
   *Upton, New York 11973, USA*

[b] *Argonne National Laboratory,*
   *Argonne, Illinois 60439, USA*

[c] *Forschungzentrum Julich GmbH,*
   *52425 Julich, Germany*

[d] *Northwestern University,*
   *Evanston, Illinois 60208, USA*
       E-mail: rumaiz@bnl.gov

ABSTRACT: We have developed a series of monolithic multi-element germanium detectors, based on sensor arrays produced by the Forschungzentrum Julich, and on Application-specific integrated circuits (ASICs) developed at Brookhaven. Devices have been made with element counts ranging from 64 to 384. These detectors are being used at NSLS-II and APS for a range of diffraction experiments, both monochromatic and energy-dispersive. Compact and powerful readout systems have been developed, based on the new generation of FPGA system-on-chip devices, which provide closely coupled multi-core processors embedded in large gate arrays. We will discuss the technical details of the systems, and present some of the results from them.

KEYWORDS: Detectors; Multielement; Germanium; Synchrotron.

# Contents

______________________________________________________________________

## 1. Ge detectors for high-energy X-rays

Germanium has been the material of choice for the detection of high-energy X- and Gamma-rays for many years due to its availability in high quality and in large volumes. Consequently, efficient detectors can be built. The high purity permits full depletion at reasonable bias voltages, and its low defect density provides excellent charge collection efficiency. Its main drawback is the need to cool the material to cryogenic temperatures to eliminate thermally-generated carriers. However, modern closed-cycle cryostats have greatly reduced that burden. Figure 1 compares the absorption properties of germanium and silicon as a function of photon energy. Typical silicon detectors are formed from high-resistivity silicon wafers of thickness 0.5 mm. For this thickness, full depletion can be achieved with bias voltages of 100V or so. As can be seen, this thickness provides reasonable efficiency (> 20%) up to energies of 25 keV. Increasing the silicon thickness to 3 mm improves the performance significantly; however such a detector would need to be biased to several kilovolts for full depletion, and would need also to be cryo-cooled to reduce its leakage current sufficiently. The advantage offered by germanium is obvious, with a 3 mm detector holding its efficiency above 20% beyond 200 keV. Our detectors described here are all 3 mm thick, and were operated at bias voltages below 300V for full depletion.

## 2. Sensor technology

The sensors used in this work were all produced at FZ Julich[1], and were based on a technology which relied on the use of trenches to provide pixel-pixel electrical isolation. This technique is simple and reliable and produces high-quality devices with excellent yield. Figure 2 shows an optical micrograph of one of these sensors. This particular one had strips with a pitch of 0.125 mm, with 30 µm wide trenches separating them. The sensor is mounted to a commercial cryo-cooler and the operating temperature of the sensor was around 100 K.

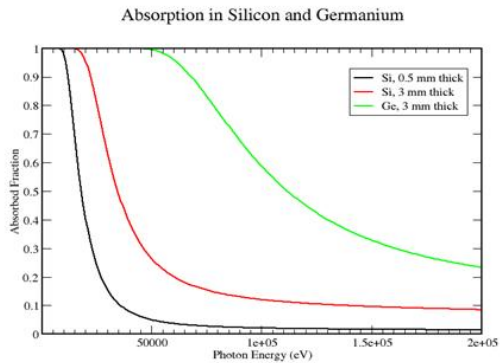

Figure 1. Comparison of absorption properties of silicon and germanium

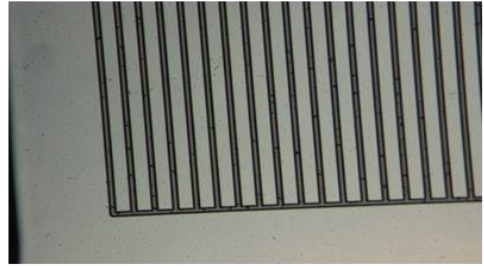

Figure 2. Optical micrograph of the sensor structure. The dark areas are the trenches, etched into the bulk germanium

## 3. Readout electronics

We initially tested these sensors using a simplified readout system based on the HERMES application specific integrated circuit (ASIC) [2]. This chip was primarily designed for photon-counting applications, but had very low noise, and the analog signals could be multiplexed to an external MCA one by one. The tests with this simple system were very encouraging, so we started a development program to fully exploit the capabilities of the sensors.

### 3.1 The MARS readout ASIC

Our first implementation of a germanium detector used a 64-strip sensor with the same readout system as the Maia detector [3]. We have since received an improved ASIC which replaces the two ASICs used in the Maia design. We will not describe the Maia system, since it has been well-described in the literature. In this section we will describe the new ASIC, MARS (Multi-element Amplifier and Readout System).

The ASIC contains 32 identical channels, each one consisting of a charge-sensitive preamplifier capable of accepting either electron or hole signals, with four programmable gains from 3600mV/fC to 600mV/fC (12.5 keV – 75 keV full scale) and a shaping amplifier with four programmable shaping times from 0.25 µs to 2 µs. Each channel also has a 2-phase peak detector and a time-to-analog converter [4]. The peak detector serves to capture the photon energy information, and also to act as temporary storage for the value until the readout system can retrieve it. The timing system can be configured in two ways. It can be configured to measure the time the shaped pulse spends over the channel threshold. We have used that system in Maia to detect pulse-pileup. For the detectors presented here we were more concerned with detecting charge-shared events, since the pixels are small compared to the detector thickness. The timing system can therefore also be configured to measure the time of arrival of a photon. Since all of the data from one ASIC is multiplexed through one differential analog port, photon events are retrieved sequentially by the data acquisition (DAQ) module. The analog timing system records the interval between the photon peak-detect signal and the actual readout of the value. Simultaneously with reading out the peak value, the DAQ system records the value of a system clock. Together, the system clock and analog time values can be used to reconstruct the real time of arrival. The data stream can then be examined for time-coincident events in neighboring strips, which is a signature of a shared event.

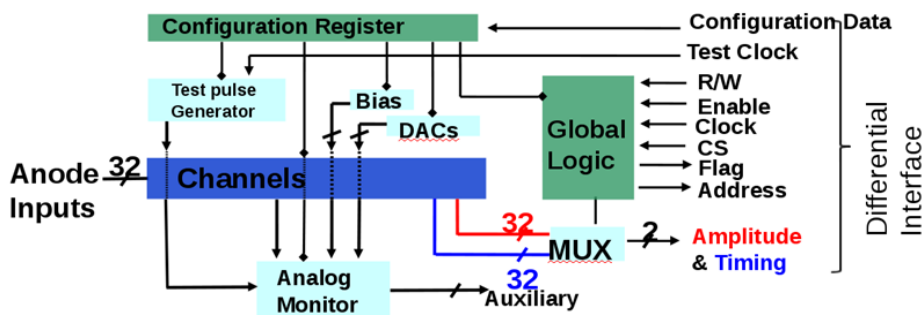

Figure 3. Block diagram of the MARS ASIC architecture.

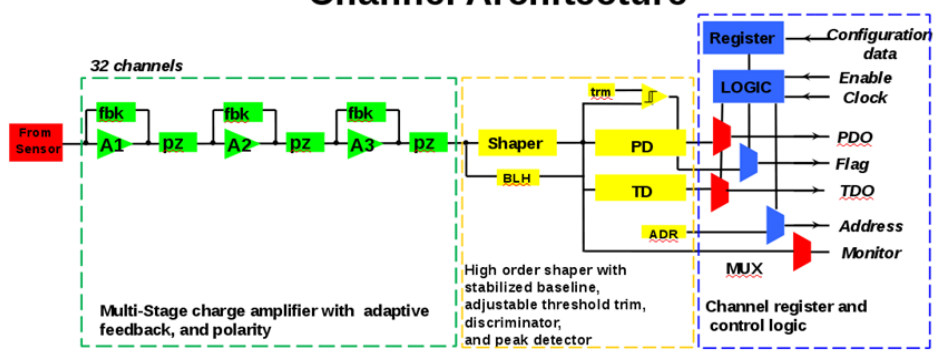

Figure 4. Architecture of a single MARS channel.

Each channel has a threshold discriminator. Pulse processing only takes place when a pulse exceeds this value. To account for channel-channel differences in discriminator offset (due to semiconductor process variations) the discriminator setting has two components: a global setting per ASIC and a per-channel trim value. In this way each channel can be trimmed to equalize actual performance. Figure 3 shows a block diagram of the ASIC architecture and figure 4 shows an individual channel.

The MARS chip was designed originally for use with silicon drift detectors. As such it was not intended to handle the larger signals expected from 200 keV photons. Since this is exactly the area where the germanium systems will operate, we designed a second version of MARS with this in mind. This second chip is called HE-MARS. This was particularly important for the energy-dispersive crystallography application, to be described later. It is identical to MARS with the exception of a reduction in the gain. HE-MARS has an energy range from 25 keV to 200 keV.

**3.2 The Ge Readout Module (GeRM)**

The interface module takes the signals from up to 12 ASICs and digitizes them and transfers them via Ethernet to a computer storage system. It has three main components:

1. The in-cryostat board housing the ASICs, mounted in close proximity to the sensor, but not thermally connected to it

2. A vacuum-feedthrough assembly which takes roughly 400 signals through the vacuum enclosure to the processing electronics.

3. A module containing a powerful field-programmable gate array (FPGA) which also has an embedded dual-core ARM processor, plus all of the components necessary to form a computer system (memory, storage, networking etc.). Figure 5 shows a block diagram of the whole system. Figures 6, 7 and 8 show realizations of the components.

The components are connected together via a pair of rigid-flex PCBs, one in-vacuum and one in air.

The readout module acts to supervise the detector operations. It runs the Linux operating system, allowing straightforward implementation of control software. Linux communicates with the FPGA firmware via memory-mapped ports, which correspond to the various functions managed by the FPGA. This type of structure leaves all of the real-time activity to the FPGA, at which it excels, and simply sets switches and registers for the FPGA to act upon. Thus the non-real-time nature of the Linux system is not an issue. To transfer data, the system has a Gigabit Ethernet connector (seen in the center-left of the photograph). It also has two SFP connectors (top left), which will allow a speed enhancement if necessary since it can accept a multi-gigabit fiber connection. One of these connectors will be used to accept the NSLS-II Event Generator signal, which in turn will allow us to synchronize our data timestamps to the NSLS-II global clock. This integrates well with the NSLS-II data acquisition strategy which will provide accurate timestamps for all data, and merge the various streams after the fact.

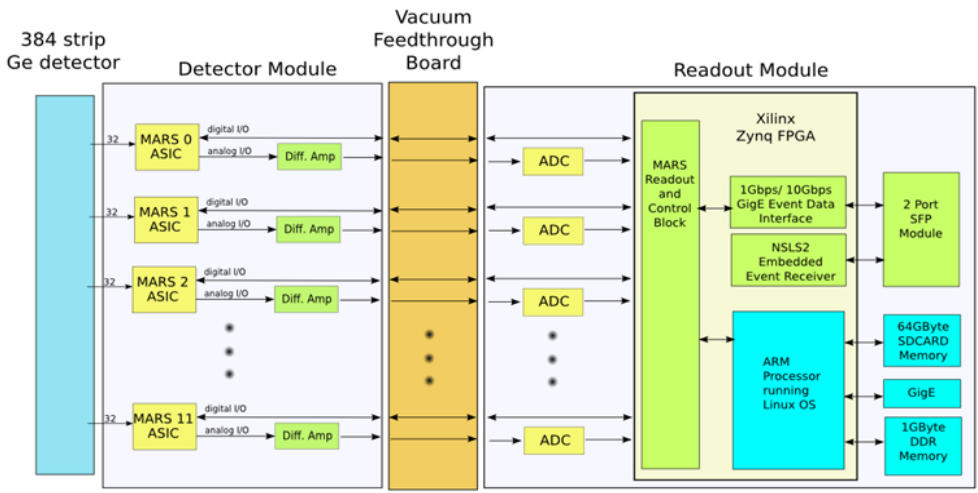

Figure 5. Block diagram of the complete detector system. It consists of three main components: A detector board with the ASICs and associated buffer amplifiers, a multi-pin vacuum feedthrough of custom design and a processor module.

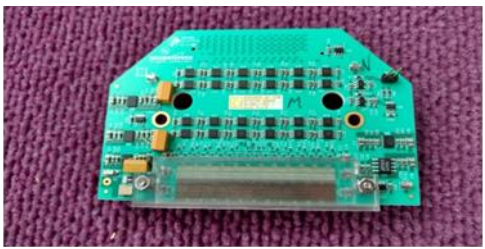

Figure 6. The detector module. It holds up to 12 ASICs plus 24 differential analog buffers to allow the 24 signals to be transferred through the vacuum vessel wall.

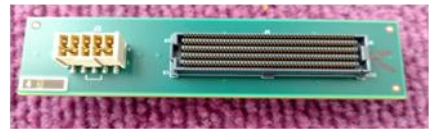

Figure 7. The multi-pin vacuum feedthrough. It can interface up to 500 signal connections plus 10 power connections. The PCB is the actual vacuum barrier, sealed by an 'O' ring.

In the following sections we will describe some of the applications of the detectors we have built.

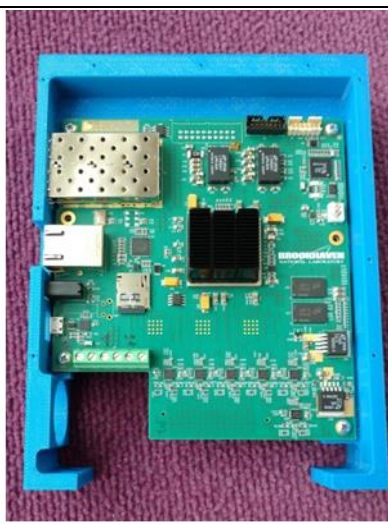

Figure 8. The Readout module. It holds an embedded computer plus large FPGA for control and readout of the detector

## 4. EDX

One of the experiments which motivated our developments was Energy Dispersive X-ray diffraction (EDX). This technique is a variation on standard Bragg diffraction for crystal structure determination. The standard experiment uses a monochromatic x-ray source and measures intensity versus diffraction angle. EDX uses a fixed diffraction angle, and measures intensity versus wavelength (i.e. energy). A typical setup is shown in figure 9a. The collimated beam and the collimated detector selects a gauge volume within the sample. Diffraction is observed only within this volume, so the crystalline structure of the entire sample can be mapped out by rastering the sample through the gauge volume in three dimensions. One of these scan dimensions can be removed using the setup in figure 9b. Here, the double-pinhole collimator is replaced by a single pinhole and a position-sensitive detector. The dimension along the beam is projected onto the detector so that its diffraction is measured simultaneously along that dimension.

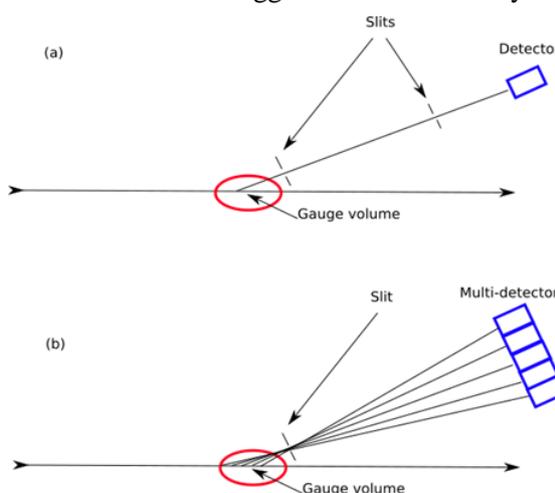

Figure 9. (a) conventional EDX setup, measures diffraction from a small gauge volume in the sample (b) multi-detector setup. Measures diffraction from all volume traversed by the incident beam.

This idea was first proposed by Dr. Zhong Zhong of NSLS-II, but was never implemented for lack of a suitable detector. This detector development has made it feasible.

### 4.1 64-strip detector

A proof of concept experiment was carried out at NSLS in 2014 with a 64-strip sensor and very simple readout electronics [2] and its success prompted us to make a more capable system for routine use at a beamline. We subsequently have built two detectors with this application in mind. The first is a 64-strip device with 0.5 mm x 8 mm strips. This was installed at BM6 at the Advanced Photon Source and used to explore the technique and its possibilities. As a test case we used samples of interest to archeologists, some human bones of medieval origin, and a synthetic bone phantom of similar size to act as a well-known object to image.

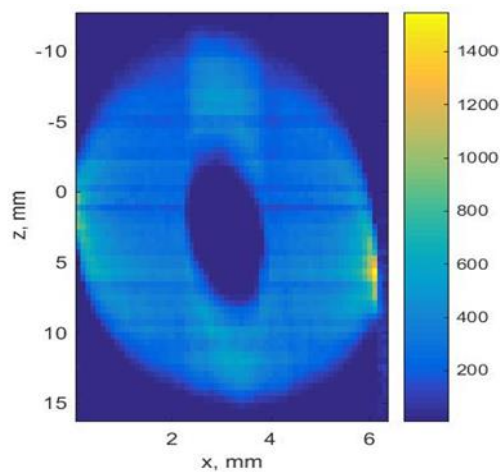

Figure 10. Tomographic slice of a bone phantom extracted from the intensities of the 2 1 .1 Bragg peak in the detector. The data is not absorption-corrected.

The advantage of this technique became very obvious. We were able to make a tomographic slice image of these samples with a single transverse scan. No rotations are necessary as is typically required for conventional tomography, and reconstruction is straightforward. This single scan can map crystal structure in two dimensions, including strains. Figure 10 shows such a tomographic slice constructed from the intensities of the 2 1 . 1 Bragg peak.

### 4.2 192-strip detector

The success of this experiment led to a request for a detector with more, smaller pixels. The 192-strip detector was the response to that request. It has 0.25 mm pitch, which is a compromise between the desire to minimize charge-sharing and to improve spatial resolution. The detector has been finished and we show initial testing results below.

At this point we already had PCBs to build the monochromatic diffraction detector, which has 384 strips of 0.125 mm pitch. To accommodate the 192-strip device, we simply omitted every other ASIC, and spread out the connecting wire-bonds to match the sensor. Figure 11 shows the assembled detector.

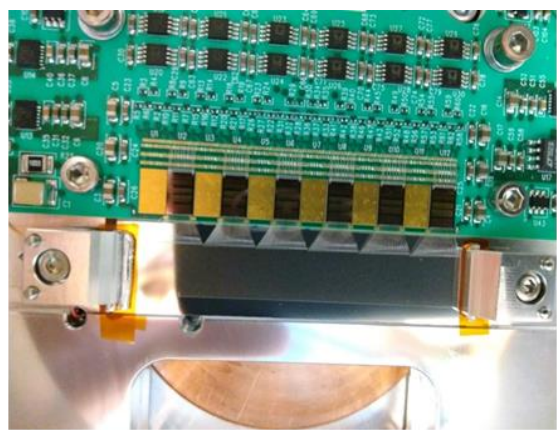

Figure 11. Photograph of the completed 192-strip EDX detector. The gold pads are vacant ASIC positions, which would be filled for the 384-strip detector. The wirebonds now need to fan out to reach the strips.

The detector has been tested using a $^{57}$Co radioactive source. This source is ideal for this detector since it has two high-energy gamma lines, one at 122 keV and the other at 136 keV, plus a 14.4 keV Mossbauer line. We used the 122 keV line and the 14.4 keV line to do a 2-point linear energy calibration for each strip. Figure 12 shows the resulting spectra, presented as a false-color image. The horizontal axis represents photon energy, and the vertical axis represents strip number. The color represents the intensity at each energy. All channels appear to be functional, having very similar performances. A Gaussian fit to the 122 keV line gave the energy resolution (FWHM) of one representative channel at the 122 keV line as 770 eV. This is to be compared with the calculated Fano-limited value of 510 eV.

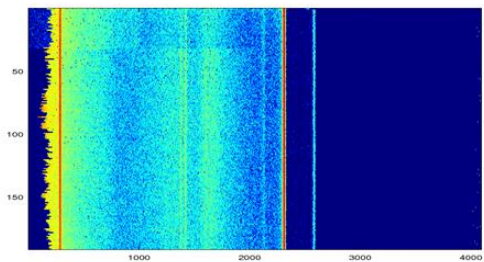

Figure 12. Channel number (vertical axis) vs energy (horizontal axis). Color indicates log(intensity) (red = high intensity, blue = zero intensity). The three strong lines are clearly seen: 14.4 keV, 122 keV and 136 keV. The weak lines around 1500 ADUs arise from the lead used to shield the source. The weak line just above 2000 ADUs is the germanium escape peak.

## 5. Monochromatic diffraction

The detector we built for this application has 384 strips, each 0.125 mm x 8 mm. The in-vacuum board is identical to the 192-strip board, but is fully-populated with 12 MARS ASICs (shown in figure 13). The remaining components of the system are also identical. We mounted the detector on the large diffractometer at the NSLS-II powder diffraction beamline, XPD, and collected data from standard samples to characterize the system. The small strip pitch leads to a significantly greater fraction of charge-shared events. The problem can be mitigated by a suitable setting of thresholds. Typically, a threshold of one-half of the full-energy value will produce optimal efficiency and uniformity of response.

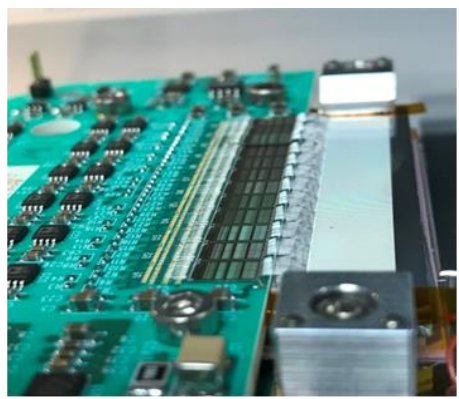

Figure 13. The 384-strip germanium detector wire-bonded to 12 MARS ASICs.

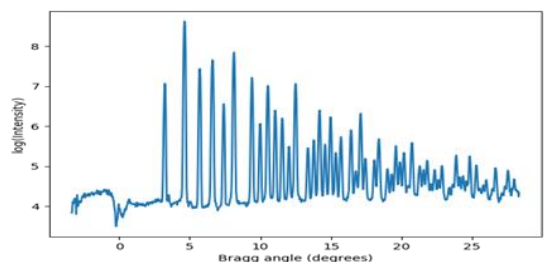

Figure 14. The powder diffraction profile from Barium Titanate, taken by the Ge strip detector at 53keV.

To test this detector we mounted it on the large 2-circle diffractometer at the NSLS-II XPD beamline. We took data from Barium Titanate at 53 keV. The threshold was raised to suppress the contribution of fluorescence in the diffraction data. Figure 14 shows a measurement over 30 degrees. We show it on a logarithmic intensity scale to bring out the quality of the weak high-angle data.

## 6. Charge sharing and position interpolation

As mentioned above, the small pitch of this sensor compared to its thickness leads to a significantly high fraction of events which are recorded by two adjacent strips, i.e. charge-shared. We have taken advantage of the MARS time-stamping capability to detect simultaneous events in neighboring strips, and recombine the fractional charges to recover the full energy of the event. Figure 15(a) shows a parametric plot of 10,000 such simultaneous 2-strip events. The diagonal lines are indicative that the events really were charge-shared, and the sum of their equivalent energies is constant. In figure 15(b) and 15(c) we compare the energy spectra before and after this reconstruction. It is clear that the peak-to-background is greatly improved when the partial events are recovered. We noticed the individual resolution to be about 700 eV whereas the reconstructed data had a resolution of about 980 eV which is consistent with expected 1.4x increase in noise. In principle, the charge sharing phenomenon can be used to provide a spatial resolution which is better than the strip pitch [5]. In the future we will attempt to implement this for our detectors.

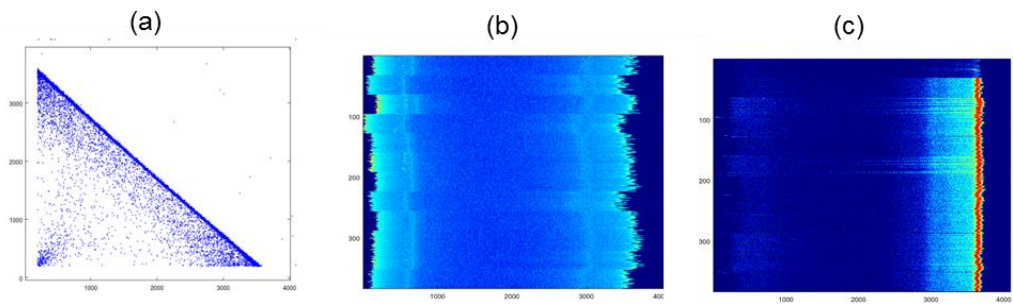

Figure 14. (a) A parametric plot of the energy detected in one strip vs the energy detected simultaneously in its neighboring strip. (b) A plot of energy (x-axis) vs strip number (y-axis). The color represents the log of the intensity. (c) the same data after energy normalization and charge-sharing reconstitution.

## 7. Summary

We have developed a toolkit allowing us to make monolithic multi-element germanium detectors for applications in synchrotron radiation research. We show preliminary results from such applications.

**Acknowledgments**